\documentclass[prl,showpacs,twocolumn,aps,superscriptaddress,a4paper]{revtex4-1}
\usepackage{dcolumn,amssymb,amsmath,amsfonts,graphicx,latexsym,color,braket}

\begin{document}

\title{Topologically driven nonequilibrium phase transitions in diagonal ensembles
}
\author{Pei Wang}
\email{wangpei@zjut.edu.cn}
\affiliation{Institute for Theoretical Physics, Georg-August-Universit\"{a}t G\"{o}ttingen,
Friedrich-Hund-Platz 1, G\"{o}ttingen 37077, Germany}
\affiliation{Department of Applied Physics, Zhejiang University of Technology, Hangzhou 310023, China}

\author{Stefan Kehrein}
\affiliation{Institute for Theoretical Physics, Georg-August-Universit\"{a}t G\"{o}ttingen,
Friedrich-Hund-Platz 1, G\"{o}ttingen 37077, Germany}

\date{\today}

\begin{abstract}
We identify a new class of topologically driven phase transitions when
calculating the Hall conductance of two-band Chern insulators in the long-time limit
after a global quench of the Hamiltonian.
The Hall conductance is expressed as the integral of the Berry curvature in the diagonal ensemble.
Even if the topological invariant of the wave function is conserved under unitary evolution,
the Hall conductance as a function of the energy gap in the post-quench Hamiltonian
displays a continuous but nonanalytic behavior,
that is it has a logarithmically divergent derivative as the gap closes.
The coefficient of this logarithmic function
is the ratio of the change of Chern number
in the ground state of the post-quench Hamiltonian to
the energy gap in the initial state.
This nonanalytic behavior is universal in two-band Chern insulators.
\end{abstract}

\maketitle

\emph{Introduction}.--
The discovery of the quantum Hall effect~\cite{klitzing,tsui}, i.e. a quantized Hall conductance
in the ground state which jumps from one plateau to another,
inspired the study of topological order~\cite{TKNN,wen}
to characterize different topological phases outside the conventional framework
of spontaneous symmetry breaking. Considerable effort has been devoted 
to understanding topological order or symmetry
protected topological (SPT) order in the ground state. More recently,
a lot of attention was devoted to the nature of topological order and SPT order
for a state driven out of equilibrium, in particular for quantum quenches
of the Hamiltonian~\cite{tsomokos09,halasz13,rahmani10,foster,fosterprb,dong14,wang14,
perfetto13,chung14}.

Consider an isolated system initially in the ground state of a Hamiltonian $\hat H_i$
and suddenly changing the Hamiltonian to $\hat H_f$. The wave function follows a unitary time
evolution, while the local observables in the long time limit settle to the prediction of
the diagonal ensemble~\cite{rigol08}, which in some cases can be reduced to a thermal ensemble or a
generalized Gibbs ensemble~\cite{rigol07,polkovnikov}.
Topological order or SPT order cannot be expressed as a local observable.
Therefore, its identification in a nonequilibrium state is far from trivial.
In the toric code model, the topological entropy
in the long time limit is found to be the same as its initial value
independent from whether the ground states of $\hat H_i$ or $\hat H_f$ are
topologically trivial or not~\cite{tsomokos09,halasz13,rahmani10}.
This result agrees with a universal argument for gapped spin liquids~\cite{chen10}. Similarly,
for the Fermi gas on a honeycomb lattice which essentially simulates the Haldane model,
the Chern number is proved to be conserved under unitary evolution~\cite{alessiol,caio15}.
However, in the two-dimensional topological superfluid,
the winding number of the retarded Green's function after a quench shows
a strong dependence on the post-quench Hamiltonian $\hat H_f$~\cite{foster,dong14},
even if the winding of the Anderson pseudo spin texture
is conserved~\cite{fosterprb}. Also in the one-dimensional
case, an analysis of tunneling spectroscopy by
coupling the system to an auxiliary thermal bath shows that the SPT order
is mostly determined by $\hat H_f$~\cite{wang14}. But in topological superconductors
with proximity-induced superconductivity, the Majorana
order parameter~\cite{perfetto13} or the entanglement spectrum~\cite{chung14} indicate
that the quenched state is topologically trivial if $\hat H_i$ and
$\hat H_f$ are in different topological phases.

To clarify the issue of SPT order far from equilibrium,
we appeal to a measurable physical quantity, namely the Hall conductance in
Chern insulators. We first study a paradigmatic
model, i.e. the Dirac model~\cite{shen}, and then extend our results
to a general two-band Chern insulator.
We find that the Chern number of the unitarily evolving wave function
is conserved and uniquely determined by $\hat H_i$.
However, while the Hall conductance
of the quenched state is a continuous
function of the energy gap in $\hat H_f$, the
derivative of this function displays a logarithmic divergence whenever the Chern number of
the ground state of $\hat H_f$ changes.
We thus identify a new class of topologically driven phase transitions
with an exotic critical behavior, which is quite different
from the orthodox one
in which the Hall conductance is discontinuous but its derivative
is zero everywhere in the phase diagram.
The discrepancy in the SPT order obtained from the Chern number 
(based on unitary time evolution) and the Hall
conductance is attributed to the fact that the latter must be calculated from the
diagonal ensemble, in which the coherence between different eigenstates of $\hat H_f$
in the wave function is lost in the long-time limit. In this experimentally relevant sense the
SPT order of quenched states depends on $\hat H_f$.

\emph{Real-time dynamics of the Chern number}.--
 The Hamiltonian of a two-band Chern insulator in two dimensions is expressed as
\begin{equation}\label{eq:twobandchernH}
 \hat H= \sum_{\vec{k}} \hat c^\dag_{\vec{k}} \mathcal{H}_{\vec{k}} \hat c_{\vec{k}},
\end{equation}
where $\hat c_{\vec{k}} = \left(\hat c_{\vec{k}1} ,\hat c_{\vec{k} 2} \right)^T$ is the fermionic
operator and $\sum_{\vec{k}}$ sums over a single Brillouin zone.
The single-particle Hamiltonian $\mathcal{H}_{\vec{k}}$ can be
decomposed into $\mathcal{H}_{\vec{k}}=\vec{d}_{\vec{k}} \cdot \vec{\sigma}$, where
$\vec{\sigma}$ denotes the Pauli matrices.

The Dirac model is a paradigm for two-band Chern insulators~\cite{shen}.
In the Dirac model,
the coefficients of the Pauli matrices are $\vec{d}_{\vec{k}}=(k_x,k_y,M-B k^2)$ with two parameters
$M$ and $B$, and $\sum_{\vec{k}}$ sums over the whole momentum plane.
The ground state is well known to be
classified by the Chern number $C=\frac{1}{2}\left( \textbf{sgn}(M)+ \textbf{sgn}(B)\right)$,
which is quantized and changes only at the phase boundary
$M=0$ or $B=0$. The Hall conductance of the ground state is simply
the Chern number in units of $e^2/h$.

At the time $t=0$, we suddenly change the Hamiltonian from $\hat H_i=\hat H (M_i,B_i)$ to $\hat H_f= \hat H(M_f,B_f)$.
Then the wave function evolves according to
$|\Psi(t)\rangle= e^{-i\hat H_f t}|\Psi(0)\rangle
= \prod_{\vec{k}} \otimes |u_{\vec{k}}(t) \rangle$,
where $|u_{\vec{k}}(t) \rangle$ is the single-particle wave function obeying
$\mathcal{H}_{\vec{k}}^f |u_{\vec{k}}(t) \rangle = i\frac{\partial}{\partial t} |u_{\vec{k}}(t) \rangle $.
The momentum is a good quantum number both in $\hat H_i$ and $\hat H_f$. Therefore, it is natural
to generalize the definition of the Chern number for the time-dependent wave function in the following way:
\begin{equation}\label{realtimechern}
 C(t) = \frac{i}{2\pi} \int d\vec{k}^2 \left( \Braket{ \frac{\partial 
 u_{\vec{k}}(t)}{\partial k_x} | \frac{\partial u_{\vec{k}}(t)}{\partial k_y}} - \text{H.c.}\right).
\end{equation}
This real-time Chern number characterizes the topological property of the wave function $|\Psi(t)\rangle$,
and can be reexpressed as
$C(t) = \frac{i}{2\pi} \int d\vec{S} \cdot \left( \triangledown_{\vec{k}} \times \vec{A}(t) \right)$,
where $\vec{S}$ denotes the $k_x$-$k_y$ plane oriented in the $k_z$-direction
and $\vec{A}(t) = \langle u_{\vec{k}}(t) | \triangledown_{\vec{k}} | u_{\vec{k}}(t)\rangle$
is the Berry connection. $C(t)$ is determined by the poles of $\vec{A}(t)$ and must remain
quantized at all times since locally deforming $\vec{A}(t)$ cannot change it.
In fact, the two poles of $\vec{A}(t)$ at $k=0$ and $k=\infty$ have conserved residues under a unitary evolution~\cite{supp},
so that for arbitrary $\hat H_i$ and $\hat H_f$ we have $C(t) \equiv C(0)$.
The Chern number of the wave function never changes although the system is driven
out of equilibrium, which agrees with the no-go theorem proved by D'Alessio and Rigol~\cite{alessiol}.
This result suggests that the SPT order of a wave function
is generally conserved after a quench if the Hamiltonian in real space
contains only local operators~\cite{chen10}.

\emph{Hall conductance in the diagonal ensemble}.--
The observation that $C(t)$ is independent of $\hat H_f$ does not imply
the absence of nonequilibrium phase transitions
because $C(t)$ is not a measurable physical quantity. In this paper, a
nonequilibrium phase transition is unambiguously indicated by the nonanalytic behavior
of observables as the post-quench Hamiltonian
$\hat H_f$ changes. We choose the Hall conductance as the indicator of nonequilibrium phase transitions.
Notice that in the ground state the Hall conductance is directly related to the Chern number.

It is well known that the Hall conductance cannot be expressed as
the expectation value of a local operator,
but must be written as the long-time response 
to an external electric field in linear response theory. This fact reflects
the topological nature of the Hall conductance and is related to
the observation that in order to measure the Hall conductance, one must couple the system to
auxiliary reservoirs.
However, coupling to reservoirs unavoidably introduces decoherence and therefore in the long-time limit the
far-from-equilibrium system will be described by the diagonal ensemble and not the unitarily evolved wave function
of the isolated system. This motivates us to pursue a definition of SPT order and topologically
driven nonequilibrium phase transitions by studying the Hall conductance in the diagonal ensemble,
which is the experimentally relevant setting.
In the long-time limit, the off-diagonal terms of the density matrix in the eigenbasis of $\hat H_f$
are averaged out~\cite{rigol08}. The time-averaged expectation value of an operator $\hat O$
can be expressed as
\begin{equation}
\begin{split}
\lim_{T\to\infty} \frac{1}{T} \int^T_0 
 dt \langle\Psi(t)| \hat O |\Psi(t)\rangle 
 & = \sum_{E}  | \langle E|\Psi(0)\rangle |^2 \langle E| \hat O |E\rangle \\
 & = \textbf{Tr} [\hat O \hat \rho],
\end{split}
 \end{equation}
where $|E\rangle$ are the eigenstates of $\hat H_f$ and $\hat \rho$ is diagonal in the basis $|E\rangle$
with the elements $| \langle E|\Psi(0)\rangle |^2$.
If the long-time limit of
$\langle\Psi(t)| \hat O |\Psi(t)\rangle$ exists, it must be determined by
$\hat \rho$, the so-called diagonal ensemble \cite{rigol08}.
While this argument is based on non-degenerate eigenenergies, the applicability of the diagonal
ensemble has also been shown in many integrable quantum many-body models \cite{rigol09,ziraldo}.

We build our formalism on the diagonal ensemble with the density matrix written as
\begin{equation}\label{densitymatrix}
 \hat \rho = \prod_{\vec{k}} \otimes
 \left( \sum_{\alpha = \pm} n_{\vec{k}\alpha} | u^f_{\vec{k}\alpha} \rangle \langle u^f_{\vec{k}\alpha} | \right),
\end{equation}
where $| u^f_{\vec{k}\alpha} \rangle$ is the eigenvector of $\mathcal{H}_{\vec{k}}^f$ and $\alpha=\pm$
denotes the upper and lower bands with the positive and negative eigenvalues $\pm |\vec{d}^f_{\vec{k}}|$, respectively.
$n_{\vec{k}\alpha}$ is the occupation number of the band $\alpha$ and can be expressed as the overlap
$n_{\vec{k}\alpha} = |\langle u^f_{\vec{k}\alpha} | u^i_{\vec{k} -} \rangle |^2$,
where $| u^i_{\vec{k} -} \rangle$ is the lower-band eigenvector of the initial Hamiltonian $\mathcal{H}_{\vec{k}}^i$,
which is in fact the initial wave function. The total occupation at each $\vec{k}$ is
conserved to be $n_{\vec{k}+}+ n_{\vec{k}-} \equiv 1$. Eq.~(\ref{densitymatrix}) is obtained by
averaging out the off-diagonal elements in $\left( | u_{\vec{k}}(t) \rangle \langle u_{\vec{k}}(t)|\right)$.

Now we calculate the Hall conductance of the diagonal ensemble in linear response theory~\cite{mahan},
i.e., we replace the equilibrium density matrix in linear response theory by the diagonal ensemble $\hat \rho$.
This replacement does not cause any problem in the formalism because
$\hat \rho$ is time-independent satisfying $[\hat \rho,\hat H_f]=0$. We can then express the Hall conductance
as the current-current correlation in the diagonal ensemble:
\begin{equation}\label{hallcond}
 \sigma_{H} = \lim_{\omega \to 0} \frac{1}{S\omega} \int^\infty_0 dt e^{i\omega t} \textbf{Tr}
 \left( \hat \rho  \left[ \hat J_y , \hat J_x(t) \right] \right),
\end{equation}
where $S$ denotes the area of the system and is conventionally set to unity.
$\hat J_{\beta}= e \sum_{\vec{k}} \hat c^\dag_{\vec{k}} \displaystyle \frac{\partial \mathcal{H}^f_{\vec{k}} }{\partial k_\beta}
\hat c_{\vec{k}}$ is the current operator along the $\beta$-direction with $e$ denoting the charge
of the particle. Following the process for obtaining the celebrated TKNN number~\cite{TKNN},
we reexpress the dimensionless Hall conductance $C_{neq} := \sigma_H/(e^2/h)$ as~\cite{supp}
\begin{equation}\label{cnessinitial}
C_{neq} = \frac{i}{2\pi} \sum_\alpha \int d\vec{k}^2 n_{\vec{k}\alpha}
\left(\Braket{
\frac{\partial u^f_{\vec{k}\alpha} }{\partial k_x } | \frac{\partial u^f_{\vec{k}\alpha} }{\partial k_y } }
- \text{H.c.} \right),
\end{equation}
which is the integral of the weighted mixture of Berry curvatures
in different bands of the post-quench Hamiltonian. In the case of $\hat H_i=\hat H_f$ (no quench),
the occupation is $n_{\vec{k}-}=1$ and $n_{{\vec{k}+}}=0$ everywhere in the Brillouin zone,
and $C_{neq}$ is just the Chern number of the initial state as we expect.
But for $\hat H_i \neq \hat H_f$,
$n_{\vec{k}\alpha} \in [0,1]$ becomes a continuous function of $\vec{k}$ so that
$C_{neq}$ is not quantized any more but can take an arbitrary value.

\begin{figure}[tbp]
\includegraphics[width=0.7\linewidth]{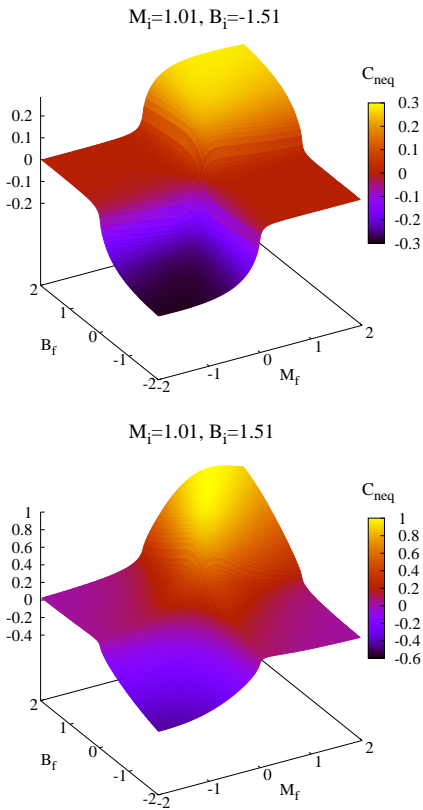}
\caption{(Color online) The Hall conductance $C_{neq}$ as a function of $(M_f,B_f)$
at different $(M_i,B_i)$ in the Dirac model. [Top panel] The initial state is
topologically trivial. [Bottom panel] The initial state is topologically nontrivial.}\label{fig:cneqfull}
\end{figure}
It is worth comparing the real time Chern number $C(t)$ in Eq.~(\ref{realtimechern})
with the dimensionless Hall conductance $C_{neq}$ in Eq.~(\ref{cnessinitial}).
The former reflects the topology
of the wave function, being quantized but not measurable, while the latter
is a true observable but not quantized.
They are both integrals of the Berry curvature, but
$C(t)$ is derived from the wave function while $C_{neq}$ follows from the diagonal ensemble
where the coherence is lost. Decoherence plays a crucial role
in understanding the SPT order of a quenched state in the long-time limit
which is a nonequilibrium steady state.

\emph{Nonanalytic behavior of Hall conductance}.--
In the Dirac model, it is straightforward to determine the Hall conductance as~\cite{supp}
\begin{equation}
 C_{neq}= \int_0^\infty d \tilde k
\frac{\left(\tilde k + (B_i \tilde k-M_i)(B_f \tilde k-M_f)\right)
\left(B_f \tilde k+M_f\right)}{4d^i_{\vec{k}} \left(d^f_{\vec{k}}\right)^4 }
\end{equation}
with $d^{i/f}_{\vec{k}}= \sqrt{\tilde k+ (B_{i/f}\tilde k-M_{i/f})^2}$.
The Hall conductance $C_{neq}$
is a function of $(M_i,B_i,M_f,B_f)$, i.e., the parameters of $\hat H_i$ and $\hat H_f$.
This function satisfies the properties:
\begin{eqnarray}\label{symm}
\begin{split}
C_{neq}(M_i,B_i,M_f,B_f) = & C_{neq}(B_i,M_i, B_f, M_f) \\   
= & -C_{neq}(-M_i,-B_i,-M_f,-B_f).
\end{split}
\end{eqnarray}
Let us study this function as $(M_f,B_f)$
changes, while $(M_i,B_i)$ is invariant, i.e., the initial state is fixed.
Due to Eq.~(\ref{symm}), we only consider the cases $M_i,B_i>0$ and $M_i>0,B_i<0$.
As shown in Fig.~\ref{fig:cneqfull}, $C_{neq}$ is a continuous function of $(M_f,B_f)$
in the whole parameter space~\cite{supp}. This result
is surprising if we consider the fact that the Chern number of the ground state
has a jump whenever $M$ or $B$ change sign.
By driving the system out of equilibrium, we smoothen the Hall conductance function.
$C_{neq}(M_f,B_f)$ has a similar shape at different $(M_i,B_i)$, reminiscent of
the function $(\textbf{sgn} (M_f)+ \textbf{sgn}(B_f))/2$, i.e., the Chern number
in the ground-state wave function of the post-quench
Hamiltonian $\hat H_f$. As $M_f,B_f\gg 0$ ($M_f,B_f\ll 0$), $C_{neq}$
takes a positive (negative) value, while $C_{neq}$ is close to zero
as $M_f$ and $B_f$ have different signs. Even if the initial state is topologically
trivial (see Fig.~\ref{fig:cneqfull}, the top panel),
the Hall conductance is finite
as $\hat H_f$ is in the nontrivial regime, but it cannot reach the quantized values $\pm e^2/h$.
When the initial state is nontrivial (see Fig.~\ref{fig:cneqfull}, the bottom panel), the Hall conductance
is suppressed as $\hat H_f$ deviates from $\hat H_i$, and can even change the
sign as $M_f$ and $B_f$ both change their signs.

While $C_{neq}(M_f,B_f)$ is continuous,
the key finding is that whenever the post-quench
Hamiltonian crosses the boundary at $M_f=0$ ($B_f=0$), the derivative of the
Hall conductance $\frac{\partial C_{neq}}{\partial M_f}$
($\frac{\partial C_{neq}}{\partial B_f}$) diverges to $+\infty$
in a logarithmic way~\cite{supp}:
\begin{equation}\label{diffdivergence}
\begin{split}
& \lim_{M_f\to 0} \frac{\partial C_{neq}}{\partial M_f} \sim \frac{-1}{2|M_i|} \ln |M_f|, \\
& \lim_{B_f\to 0} \frac{\partial C_{neq}}{\partial B_f} \sim \frac{-1}{2|B_i|} \ln |B_f|.
\end{split}
\end{equation}
As $M_f\to 0$, $\frac{\partial C_{neq}}{\partial M_f}$ as a function of $\left(\ln |M_f|\right)$
asymptotically approaches a
straight line with the slope $-1/(2|M_i|)$, which is independent of $B_f, B_i$ and the side from which
$M_f$ goes to zero. As $B_f\to 0$, $\frac{\partial C_{neq}}{\partial B_f}$
has a similar divergence since $C_{neq}$ is invariant under the exchange of $M_f$ and $B_f$
according to Eq.~(\ref{symm}).
We identify a nonequilibrium phase transition when the Chern number
in the ground state of
$\hat H_f$ changes.
The critical behavior of this phase transition
is exotic, compared to that of ground-state phase transitions in which the Hall conductance
has a zero derivative everywhere but
displays a discontinuity at the phase boundary.

This phase transition reveals different nonequilibrium phases which
share the common symmetries of the Dirac model. Apparently,
the broken symmetry picture does not account for this transition,
which must be topologically driven. Interestingly,
the topological invariant of the wave function $C(t)$
is independent of $(M_f,B_f)$, and then fails to
characterize different phases in this nonequilibrium phase transition.
One can assign the Chern number $C(\hat H_f)$ of the ground-state wave function of $\hat H_f$
to each nonequilibrium phase to distinguish them.
We will see that the change of $C(\hat H_f)$
determines the character of this nonequilibrium
phase transition in a general two-band Chern insulator.

Now let us consider a general two-band Chern insulator in two dimensions
with the Hamiltonian given by Eq.~(\ref{eq:twobandchernH}).
The coefficient vector
$\vec{d}_{\vec{k}}=\left( d_{1\vec{k}},d_{2\vec{k}},d_{3\vec{k}}\right)$
is different from model to model. But the nonanalytic behavior of Hall conductance
is insensitive to the change of $\vec{d}_{\vec{k}}$. Instead,
it depends only upon the lowest-order expansion of $\vec{d}_{\vec{k}}$
at the momentum
$\vec{q}$ where the energy gap closes ($d_{\vec{q}}=0$) at a phase boundary.
In a generic model, two components of $\vec{d}_{\vec{q}}$
must be zero. Let us suppose them to be $d_{1\vec{q}}$ and $d_{2\vec{q}}$
without loss of generality.
The energy gap is $d_{\vec{q}}=|d_{3\vec{q}}|$
when the system is close to the phase boundary.
$d_{3\vec{q}}$ is a free parameter in the
Hamiltonian (the gap parameter), which is denoted by $m$.
Note that $m=M$ in the Dirac model.

Suppose that the system is initially in a ground state
with the gap parameter $m=m_i$, before we suddenly
change $m$ in the Hamiltonian from $m_i$ to $m_f$.
We measure the Hall conductance in the long time limit.
The Hall conductance $C_{neq}$ is a function
of $m_f$, while we fix $m_i$ to be nonzero.
The function $C_{neq}(m_f)$ is continuous but nonanalytic at $m_f=0$,
where the gap of the post-quench Hamiltonian $\hat H_f$ closes.
The derivative of $C_{neq}(m_f)$ satisfies~\cite{supp}
\begin{equation}\label{eq:central}
 \lim_{m_f\to 0} \frac{dC_{neq}}{dm_f} \sim
 \frac{\displaystyle\lim_{m_f\to 0^-} C(m_f)- \displaystyle\lim_{m_f\to 0^+} C(m_f) }{2|m_i|}
 \ln |m_f|,
\end{equation}
where $C(m_f)$ denotes the Chern number in the ground-state wave function
of $\hat H_f$.
The derivative of the Hall conductance with respect to
the energy gap in $\hat H_f$ is logarithmically divergent as the gap closes.
And the coefficient of this logarithmic function
is the ratio of the change of Chern number
in the ground state of $\hat H_f$
to the energy gap in the initial state.
Eq.~(\ref{eq:central}) relates the nonequilibrium phase
transition in quenched states to the topological
phase transition in ground states, indicating that
this nonequilibrium phase transition is in fact topologically
driven. Eq.~(\ref{diffdivergence}) for the Dirac model is
a special case of Eq.~(\ref{eq:central}) as the
change of Chern number is $-1$.



\emph{Conclusions}.--
In summary, we find a new class of topologically driven phase transitions in
quenched states of two-band Chern insulators, which are characterized by
the Hall conductance as a continuous function of the energy gap in
the post-quench Hamiltonian $\hat H_f$
with a logarithmically divergent derivative.
The asymptotic behavior of the Hall conductance function
is determined by the ratio of the change of Chern number in the ground state of $\hat H_f$
to the energy gap in the initial state,
which is universal in two-band Chern insulators.
We obtain the Hall conductance by applying linear response theory
in the diagonal ensemble of the system, which is the physically correct description of the 
long-time limit in a far-from-equilibrium quench setup.
The topological
invariant of the real-time wave function fails to predict this phase transition,
which can only be correctly identified in the diagonal ensemble where decoherence effects are taken into account.
Our finding indicates the possibility
of exotic topological phase transitions in systems far from equilibrium.

Finally, we discuss the conditions for observing
this phase transition in experiments. The nonequilibrium distribution
of particles is responsible for the logarithmically divergent
derivative of the Hall conductance.
Ultracold atomic gases are known to be
well isolated from the environment and suitable
for studying the quench dynamics of many-body quantum systems~\cite{greiner}.
The Haldane model~\cite{haldane}
was recently realized with cold atoms in an optical lattice~\cite{jotzu14}.
The Haldane model is a two-band Chern insulator, in which
the quenched-state Hall conductance displays the
nonanalytic behavior in Eq.~(\ref{eq:central})~\cite{universalargue}.
The measurement of conductances in cold atoms
is difficult, but a two-terminal setup was implemented recently~\cite{brantut12,stadler12}.
We expect that our prediction can be checked
in a four-terminal setup made of cold atoms simulating the Haldane model.


\emph{Acknowledgement}.--
We thank Prof. Q. Niu for inspiring discussions. We thank J. Oberreuter
and M. Medvedyeva for their help in preparing the paper. Pei Wang is supported by
NSFC under Grant No.~11304280, and by China Scholarship Council.
S.~K. was supported through SFB 1073 (project B03) of the Deutsche Forschungsgemeinschaft (DFG).

\clearpage

\appendix

\section*{Supplementary material}

\section{The real-time Chern number $C(t)$ in the Dirac model}
\label{app:realtime}

We express the real-time Chern number as
\begin{equation}
 C(t) = \frac{i}{2\pi} \int d\vec{S} \cdot \left( \triangledown_{\vec{k}} \times \vec{A}(t) \right),
\end{equation}
where $\vec{A}(t) = \langle u_{\vec{k}}(t) | \triangledown_{\vec{k}} | u_{\vec{k}}(t)\rangle$
is the Berry connection
with $|u_{\vec{k}}(t)\rangle = (\phi_1,\phi_2)^T$ denoting the
single-particle wave function.
In the Dirac model, it is straightforward to calculate the wave function and obtain
\begin{equation}
\begin{split}
\phi_1(t) = & \frac{1}{\sqrt{2d^i_{\vec{k}}(d^i_{\vec{k}}-
d^i_{3\vec{k}})}} \big[ -\cos(d^f_{\vec{k}} t)(d^i_{\vec{k}}-d^i_{3\vec{k}}) \\ &
 + i\sin(d^f_{\vec{k}}t) \frac{d^f_{3\vec{k}}\left(d^i_{\vec{k}}-d^i_{3\vec{k}}\right)
 -k^2}{d^f_{\vec{k}}}  \big],
\end{split}
\end{equation}
and
\begin{equation}
\begin{split}
\phi_2(t) = & \frac{k_+}{2d^f_{\vec{k}}\sqrt{2d^i_{\vec{k}}(d^i_{\vec{k}}
-d^i_{3\vec{k}})}} \big[ 2d^f_{\vec{k}}\cos(d^f_{\vec{k}} t) \\ &
+ 2i\sin(d^f_{\vec{k}}t) (d^i_{\vec{k}}-d^i_{3\vec{k}}+d^f_{3\vec{k}}) \big],
\end{split}
\end{equation}
where $k_+ = k_x+ik_y$, $\vec{d}^{i/f}_{\vec{k}}=(d^{i/f}_{1\vec{k}},d^{i/f}_{2\vec{k}},
d^{i/f}_{3\vec{k}})$ is the coefficient vector in the initial and post-quench Hamiltonians,
respectively, and ${d}^{i/f}_{\vec{k}}$ is the length of $\vec{d}^{i/f}_{\vec{k}}$.
We divide the Berry connection into $\vec{A}(t)=\vec{A}_1(t)+\vec{A}_2(t)$
with $\vec{A}_\alpha=\phi^*_\alpha \triangledown_{\vec{k}} \phi_\alpha$.
Noticing that $\phi_1$ is a function of $k=\sqrt{k_x^2+k_y^2}$, we
immediately know that $\left( \triangledown_{\vec{k}} \times \vec{A}_1 \right)$
must be zero, so that $\vec{A}_1$ does not contribute to $C(t)$.
We again divide $\vec{A}_2$ into the irrelevant term $\vec{A}_{2a}$ with a zero curl and
the relevant term $\vec{A}_{2b}$ with its imaginary part written as
\begin{equation}
\begin{split}
\textbf{Im}[ \vec{A}_{2b} ] = & \left[ \cos^2(d^f_{\vec{k}} t)
+ \sin^2(d^f_{\vec{k}} t) \frac{(d^i_{\vec{k}}-d^i_{3\vec{k}}+d^f_{3\vec{k}})^2}
{\left(d^f_{\vec{k}}\right)^2} \right] \\
& \times \frac{-k_y \vec{x} + k_x \vec{y}}{2d^i_{\vec{k}}(d^i_{\vec{k}}-d^i_{3\vec{k}})},
\end{split}
\end{equation}
where $\vec{x}$ and $\vec{y}$ denote the unit vectors
in the momentum plane.

Now we reexpress the Chern number by the vector field $\textbf{Im}[ \vec{A}_{2b} ]$ as
\begin{equation}
  C(t) = \frac{-1}{2\pi} \int d\vec{S} \cdot \left( \triangledown_{\vec{k}} \times \textbf{Im}[\vec{A}_{2b}(t)] \right).
\end{equation}
$\textbf{Im}[ \vec{A}_{2b} ]$
is a vortex field with two poles at zero and infinity, respectively.
Applying the Kelvin-Stokes theorem in an annulus with inner radius $r$ and
outer radius $R$, and then taking the limit $r\to 0$ and $R\to \infty$, we obtain
\begin{equation}
\begin{split}
-2\pi C(t) =&  \lim_{r\to 0,R\to \infty} \int_{r\leq k\leq R} 
  d\vec{S} \cdot \left( \triangledown_{\vec{k}} \times \textbf{Im}[\vec{A}_{2b}] \right) \\
  = & \left( \lim_{R\to \infty} \oint_{k=R}
  - \lim_{r\to 0} \oint_{k=r} \right) \textbf{Im}[\vec{A}_{2b}] \cdot d\vec{k} \\
  = & \lim_{R\to \infty} \left( 2\pi R|\textbf{Im}[\vec{A}_{2b}]|_{k=R} \right) \\ & -  
 \lim_{r\to 0} \left( 2\pi r  |\textbf{Im}[\vec{A}_{2b}]|_{k=r}  \right),
\end{split} 
  \end{equation}
where $|\textbf{Im}[\vec{A}_{2b}]|_{k=R}$ denotes the length of the vector
$\textbf{Im}[\vec{A}_{2b}]$ at the circle of radius $\left(k=R\right)$. The first limit
evaluates $\big(\pi(1-\textbf{sgn}(B_i))\big)$, while
the second limit evaluates $\big(\pi(1+\textbf{sgn}(M_i))\big)$, being both
time-independent. In other words, the residues of
$\textbf{Im}[\vec{A}_{2b}]$ at zero
and infinity are both time-invariant,
which leads to a conserved Chern number:
\begin{equation}
 C(t) \equiv \frac{1}{2} \left( \textbf{sgn}(M_i) + \textbf{sgn}(B_i) \right). 
\end{equation}

\section{The Hall conductance of quenched states}
\label{app:hall}

In this section, we first show how to express the Hall conductance of quenched states as the
integral of the Berry curvature. Our derivation is a straightforward
extension of the work by Thouless {\it et al.}~\cite{TKNN:app}.
We then express the Hall conductance by using the
coefficient vectors in two-band Chern insulators.

In linear response theory, the Hall conductance is written as
\begin{equation}\label{hallcond:app}
 \sigma_{H} = \lim_{\omega \to 0} \frac{1}{S\omega} \int^\infty_0 dt e^{i\omega t-\eta |t|} \textbf{Tr}
 \left( \hat \rho  \left[ \hat J_y , \hat J_x(t) \right] \right),
\end{equation}
where $\eta$ is an infinitesimal number corresponding to the adiabatic switch-on
of an external electric field, and $\omega$ is the frequency of
the electric field with
the limit ${\omega\to 0}$ corresponding to the dc-conductance.
The diagonal ensemble is known to be
$\hat \rho = \prod_{\vec{k}} \otimes \left( \sum_{\alpha} n_{\vec{k}\alpha}
| u^f_{\vec{k}\alpha} \rangle \langle u^f_{\vec{k}\alpha} | \right)$,
which is a product state. Due to
the conversation law $\sum_\alpha n_{\vec{k}\alpha} =1 $,
the state of the system is limited
in a subspace of the Fock space in which the empty or doubly-occupied states
at each momentum are excluded.
We can then reexpress $\sigma_H$ in the first-quantization language as
\begin{equation}
\begin{split}
 \sigma_{H} = \lim_{\omega \to 0} \frac{1}{S\omega} \int^\infty_0 & dt e^{i\omega t-\eta |t|}\\ & \times
\sum_{\vec{k},\alpha} n_{\vec{k}\alpha} \Bra{ u^f_{\vec{k}\alpha}} \left[ \hat J^y_{\vec{k}} , \hat J^x_{\vec{k}} (t) \right]
\Ket{u^f_{\vec{k}\alpha}},
\end{split}
\end{equation}
where the momentum-resolved current operator is
$\hat J^{\beta}_{\vec{k}}= e \displaystyle \frac{\partial \mathcal{H}^f_{\vec{k}} }{\partial k_\beta}$.
Since we are interested in the dc Hall conductance which is a real number,
we take the real part of $\sigma_H$ and obtain
\begin{equation}
\begin{split}
 \textbf{Re}\sigma_H = & \frac{\sigma_H + \sigma_H^*}{2} \\
 = & \frac{-i e^2}{S} \sum_{\vec{k},\alpha,\beta}\frac{n_{\vec{k}\alpha}}{(\epsilon_{\vec{k}\alpha}
 -\epsilon_{\vec{k}\beta})^2} \\ & \times \biggl[ \Bra{ u^f_{\vec{k}\alpha}}
 \frac{\partial \mathcal{H}^f_{\vec{k}} }{\partial k_y} \Ket{u^f_{\vec{k}\beta}}
 \Bra{ u^f_{\vec{k}\beta} }
 \frac{\partial \mathcal{H}^f_{\vec{k}} }{\partial k_x} \Ket{u^f_{\vec{k}\alpha}} - \text{H.c.} \biggr],
 \end{split}
\end{equation}
where $\epsilon_{\vec{k}\alpha}$ denotes the eigenvalue of $\mathcal{H}^f_{\vec{k}}$.
We make use of the relation $\mathcal{H}^f_{\vec{k}} = \sum_\alpha \epsilon_{\vec{k}\alpha}
| u^f_{\vec{k}\alpha} \rangle\langle u^f_{\vec{k}\alpha} |$ and finally obtain
\begin{equation}\label{supp:cnessinitial}
\begin{split}
C_{neq} = & \frac{\textbf{Re}\sigma_H }{e^2/h} \\
= & \frac{i}{2\pi} \sum_\alpha \int d\vec{k}^2 n_{\vec{k}\alpha} \left(\Braket{
\frac{\partial u^f_{\vec{k}\alpha} }{\partial k_x } | \frac{\partial u^f_{\vec{k}\alpha} }{\partial k_y } } 
- \text{H.c.} \right).
\end{split}
\end{equation}

In a two-band Chern insulator with the Hamiltonian
$\mathcal{H}_{\vec{k}}=\vec{d}_{\vec{k}}\cdot \vec{\sigma}$,
the Berry curvatures in different bands
are opposite to each other. By using this and the conservation law $n_{\vec{k}+}+ n_{\vec{k}-} \equiv 1$,
we reexpress Eq.~(\ref{supp:cnessinitial}) as
\begin{equation}\label{eq:hallexptwo}
C_{neq} = \int d\vec{k}^2 \cos \theta \cdot \mathcal{C},
\end{equation}
where $\mathcal{C}$ denotes the Berry curvature in the lower-band of
the post-quench Hamiltonian $\hat H_f$
and can be expressed as
\begin{equation}\label{eq:twobandberryc}
\mathcal{C} = \frac{\left( \displaystyle\frac{\partial \vec{d}^f_{\vec{k}}}{\partial k_x}\times
 \frac{\partial \vec{d}^f_{\vec{k}}}{\partial k_y}\right) \cdot \vec{d}^f_{\vec{k}}}
 {4\pi (d^f_{\vec{k}})^3},
\end{equation}
and $\cos \theta$ is the occupation factor defined as
\begin{equation}
 \begin{split}
  \cos \theta := & (2n_{\vec{k}-}-1) \\
  =& (\vec{d}^f_{\vec{k}} \cdot \vec{d}^i_{\vec{k}})/({d}^f_{\vec{k}} {d}^i_{\vec{k}})
 \end{split}
\end{equation}
with $\theta$ denoting the angle between $\vec{d}^i_{\vec{k}}$ and $\vec{d}^f_{\vec{k}}$.
$\vec{d}^i_{\vec{k}}$ and $\vec{d}^f_{\vec{k}}$ are the coefficients
of the Pauli matrices in the initial and post-quench Hamiltonians, respectively, and
${d}^i_{\vec{k}}$ and ${d}^f_{\vec{k}}$ are the length of $\vec{d}^i_{\vec{k}}$ and $\vec{d}^f_{\vec{k}}$,
respectively.

\section{Continuity and nonanalyticity of the Hall conductance
in the Dirac model}

In this section, we show how to prove the continuity of $C_{neq}(M_f,B_f)$
and the logarithmic divergence of its derivative
at the phase boundary. We only prove the case at $M_f= 0$ when
$B_f$ is fixed to be nonzero, since
$C_{neq}(M_f,B_f)$ is invariant under the exchange of $M_f$ and $B_f$.

In the Dirac model, both $\mathcal{C}$ and $\cos \theta$ are
rotationally invariant in the $k_x$-$k_y$ plane. We can then
carry out the azimuthal integration in Eq.~(\ref{eq:hallexptwo}).
By making a substitution $\tilde k=k^2$, we express the Hall conductance as
\begin{equation}
 C_{neq}= \int_0^\infty d \tilde k \cos \theta \cdot \mathcal{C},
\end{equation}
where the Berry curvature is
\begin{equation}
 \mathcal{C} = \frac{1}{4} \frac{B_f \tilde k+M_f}{\left(d^f_{\vec{k}}\right)^3},
\end{equation}
and the occupation factor is
\begin{equation}
 \cos \theta = \frac{\tilde k + (B_i \tilde k-M_i)(B_f \tilde k-M_f)}{d^i_{\vec{k}} d^f_{\vec{k}}}
\end{equation}
with $d^{i/f}_{\vec{k}}= \sqrt{\tilde k+ (B_{i/f}\tilde k-M_{i/f})^2}$.
At $M_f \neq 0$, we can express the derivative of $C_{neq}$ as
\begin{equation}
\frac{\partial{C_{neq}}}{\partial M_f}= \int_0^\infty d \tilde k \frac{ \partial (\cos \theta \cdot \mathcal{C})}{\partial M_f}.
\end{equation}
A straightforward observation is that both $\cos \theta (\tilde k)$
and $\mathcal{C}(\tilde k)$ are smooth functions for $\tilde k \in (0,\infty)$.
However, they
do not uniformly converge to $(\cos \theta)_{M_f=0}$ or $\mathcal{C}_{M_f=0}$ as $M_f\to 0$.
The unique singularity is $\tilde k =0$, at which we have
$\lim_{\tilde k \to 0} \lim_{M_f\to 0}\cos \theta =0 $ but
$\lim_{M_f\to 0} \lim_{\tilde k \to 0} \cos \theta = \textbf{sgn}(M_i)\textbf{sgn}(M_f)$.
And $\mathcal{C}\left(\tilde k=0\right)= \textbf{sgn}(M_f)/(4M_f^2)$ is divergent as $M_f \to 0$.
We divide the integral into two parts: $\int_0^\infty d\tilde k = \int_0^\eta d\tilde k
+ \int_\eta^\infty d\tilde k$ with $\eta>0$
a number that can be arbitrarily small. The second integral is
a smooth function of $M_f$, which can be proved
by studying the asymptotic behavior of
$(\cos \theta \cdot \mathcal{C})$ in the limit $\tilde k\to \infty$, or more precisely,
by making a substitution $\tilde k \to 1/\tilde k$ in the integral.
In fact, $\tilde k = \infty$ is a true singularity at the boundary $B_f= 0$,
where $\tilde k =0$ is a regular point, since $\cos \theta$ and $\mathcal{C}$
are invariant under the substitution $\tilde k \leftrightarrow 1/\tilde k$
and $M_{i/f} \leftrightarrow B_{i/f}$. If there is any nonanalytic behavior
in the function $C_{neq}(M_f)$,
it must come from the first integral denoted by $C^\eta_{neq}$ next.
Interestingly, we can choose an arbitrarily small $\eta$ so that $d^i_{\vec{k}}$
in $\cos \theta$ converges to a constant $d^i_{\vec{k}}=|M_i|$. We then obtain
\begin{equation}
 C^\eta_{neq}=\int^\eta_0 d\tilde k \frac{\left(\tilde k + (B_i \tilde k-M_i)(B_f \tilde k-M_f)\right)(B_f \tilde k+M_f)}
 {4|M_i| \left( \tilde k+ (B_{f}\tilde k-M_{f})^2\right)^2}.
\end{equation}
The calculation of this integral is straightforward since the integrand
is rational.

We express the result as $C^\eta_{neq}= F(\eta)-F(0)$ with $F$ denoting
the original function. The expression of $F$ is lengthy, but it is an elementary function.
$F(\eta)$ is a smooth function of $M_f$, while $F(0)$ is expressed as
\begin{widetext}
\begin{equation}\label{fzeroexp}
 \begin{split}
  F(0)= \frac{1}{8|M_i| B^2_f }& \biggl[ \frac{2B_f(B_i+B_f)M_f -B_i }{\sqrt{1-4B_fM_f}} 
  \ln \frac{\left(2B_f^2M_f^2-4B_fM_f+1\right)\sqrt{1-4B_fM_f} - 8B_f^2M_f^2 +6B_fM_f -1}{M_f^2} \\ & 
  + B_i \ln M_f^2+2B_i-2B_f \biggr].
 \end{split}
\end{equation}
\end{widetext}
We are interested in $F(0)$
as a function of $M_f$ in the neighborhood of the phase boundary $M_f=0$.
We notice that $\sqrt{1-4B_fM_f}$ can be expanded at $M_f=0$ into
\begin{equation}
\begin{split}
 \sqrt{1-4B_fM_f} = & 1-2B_fM_f-2B_f^2M_f^2-4B_f^3M_f^3 \\ & 
 -10B_f^4M_f^4 + \mathcal{O} (M_f^5).
\end{split}
 \end{equation}
We substitute this expression into Eq.~(\ref{fzeroexp}) and obtain
\begin{equation}
 \begin{split}
  F(0)= & \frac{2B_i-2B_f-B_i \ln(2B_f^4)}{8|M_i|B_f^2} + \frac{M_f}{4|M_i|}\ln M_f^2 \\ & + \mathcal{O}(M_f) 
  - \frac{B_i}{8|M_i|B_f^2}
  \ln\left( 1+\mathcal{O}(M_f)\right) \\ &  + \mathcal{O}(M_f) \ln \left( 1+\mathcal{O}(M_f)\right) 
  + \mathcal{O}(M_f^2) \ln | M_f| .
 \end{split}
\end{equation}
The first term is independent of $M_f$. The second term
is a continuous function of $M_f$,
but its derivative with respect to $M_f$ is divergent as $M_f \to 0$.
All the other terms are continuous functions of $M_f$,
and their derivatives with respect to $M_f$ are finite at $M_f=0$.
The asymptotic behavior of $\partial C_{neq}/\partial M_f$
is uniquely determined by the second term. The function
$C_{neq}(M_f)$ then asymptotically approaches $\left(-M_f \ln |M_f|/(2|M_i|) + const.\right)$
in the limit $M_f \to 0$.
This immediately leads to our results that
$C_{neq}(M_f)$ is continuous~\cite{continuity} and
$\partial C_{neq}/\partial M_f$ is logarithmically divergent as
\begin{equation}
 \lim_{M_f\to 0} \frac{\partial C_{neq}}{\partial M_f} \sim \frac{-1}{2|M_i|} \ln |M_f|.
\end{equation}

Furthermore, we calculate the Hall conductance by numerically integrate
Eq.~(\ref{eq:hallexptwo}).
\begin{figure}[tbp]
\includegraphics[width=1.0\linewidth]{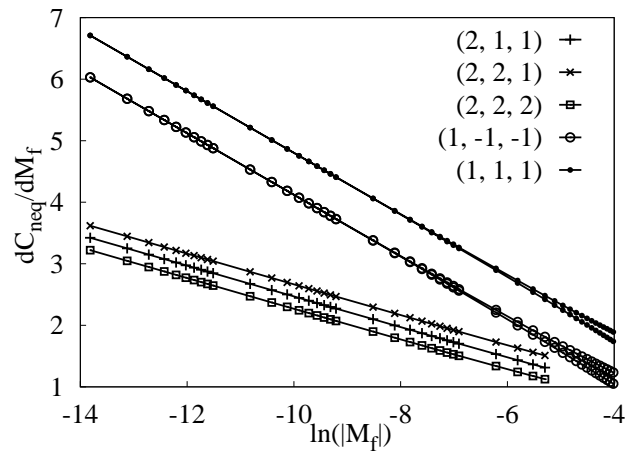}
\caption{${\partial C_{neq}}/{\partial M_f}$ as a function of $(\ln |M_f|)$ at different $(M_i,B_i,B_f)$
in the Dirac model.
Note the curves at $M_i=1$, in which we simultaneously plot the data at $M_f\to 0^+$
and at $M_f \to 0^-$,
which are in fact undistinguishable at small $|M_f|$.}\label{fig:diffcneq}
\end{figure}
We plot $\frac{\partial C_{neq}}{\partial M_f}$ as a function of $\ln |M_f|$ in Fig.~\ref{fig:diffcneq}.
In the limit $M_f\to 0$, the curves asymptotically approach straight lines
with the slope $-1/2|M_i|$, which is independent of
$B_i$, $B_f$ and the side from which $M_f$ goes to zero.
The numerical result coincides well with our analysis.

\section{Universal nonanalytic behavior of
the Hall conductance function in two-band Chern insulators}

Let us consider a general two-band Chern insulator in two dimensions
with the Hamiltonian expressed as
\begin{equation}\label{eq:supp:twobandchernH}
 \hat H= \sum_{\vec{k}} \hat c^\dag_{\vec{k}} \mathcal{H}_{\vec{k}} \hat c_{\vec{k}},
\end{equation}
where the single-particle Hamiltonian can be decomposed into
$\mathcal{H}_{\vec{k}}=\vec{d}_{\vec{k}} \cdot \vec{\sigma}$ with
$\vec{\sigma}$ denoting the Pauli matrices.
Examples include
the Dirac model, the Haldane model~\cite{supp:haldane} or the Kitaev honeycomb model
in the fermionic basis~\cite{kitaev06,chen08}.
The coefficient vector
$\vec{d}_{\vec{k}}=\left( d_{1\vec{k}},d_{2\vec{k}},d_{3\vec{k}}\right)$
is different from model to model. But the nonanalytic behavior of Hall conductance
is insensitive to the change of $\vec{d}_{\vec{k}}$,
but depends only upon the lowest-order expansion of $\vec{d}_{\vec{k}}$
at the singularities of the Berry curvature.

Let us first show how the Chern number of the ground-state wave function
is related to the expansion of $\vec{d}_{\vec{k}}$. The Chern number
is expressed by the Berry connection as
\begin{equation}
 C = \frac{-1}{2\pi} \int d\vec{S} \cdot \left( \triangledown_{\vec{k}} \times \textbf{Im} \vec{A} \right)
\end{equation}
with
\begin{equation}
\textbf{Im} \vec{A} = \frac{d_{1\vec{k}}
\triangledown_{\vec{k}} d_{2\vec{k}}-d_{2\vec{k}}\triangledown_{\vec{k}} d_{1\vec{k}}}
 {2d_{\vec{k}} (d_{\vec{k}}-d_{3\vec{k}})}.
\end{equation}
The Chern number must be zero when $\vec{A}$ has no singularity in the Brillouin zone.
According to Kelvin-Stokes theorem,
the Chern number can be expressed as the line integral of $\vec{A}$ over
the boundaries of the infinitesimal neighborhoods of singularities.
Suppose that $\vec{A}$ has a set of singularities
$\vec{q}_1, \vec{q}_2, \cdots, \vec{q}_N$ in a single Brillouin zone.
The Chern number can be expressed as
\begin{equation}
 C=\sum_{j=1}^N C^{(\vec{q}_j)}
\end{equation}
with
\begin{equation}\label{eq:chernnumberA}
C^{(\vec{q}_j)}
=\frac{1}{2\pi} \lim_{\eta\to 0} \oint_{\partial B_\eta(\vec{q}_j)} \textbf{Im}\vec{A}\cdot d\vec{k},
\end{equation}
where $\partial B_\eta(\vec{q}_j)$ denotes the boundary of
a circle of radius $\sqrt{\eta}$ centered at $\vec{q}_j$, and the
integral is along the anticlockwise direction. Here we
do not consider the singularity at infinity,
since the Brillouin zone is finite in a generic model.

In general, a singularity of $\vec{A}$ is a point $\vec{q}$ at which
$\left(d_{\vec{q}}-d_{3\vec{q}}=0 \right)$ and then
$d_{1\vec{q}}=d_{2\vec{q}}=0$. In a generic model,
$d_{\vec{q}}=|d_{3\vec{q}}|$ is the energy gap
when the system is close to the phase boundary.
$d_{3\vec{q}}$ is a free parameter in the
Hamiltonian, which is denoted by $m$ next.
Note that $m=M$ in the Dirac model.
$m$ is zero if and only if
the energy gap closes accompanied by a change of the Chern number.

The Berry connection can be reexpressed as
\begin{equation}
 \textbf{Im} \vec{A}= \left(\frac{d_{\vec{k}}+d_{3\vec{k}}}{2d_{\vec{k}} }\right)
 \left(\frac{d_{1\vec{k}}\triangledown_{\vec{k}} d_{2\vec{k}}-d_{2\vec{k}}
 \triangledown_{\vec{k}} d_{1\vec{k}}}
 {\left(d_{1\vec{k}}\right)^2+\left(d_{2\vec{k}}\right)^2}\right).
\end{equation}
Since $d_{\vec{q}}$ and $d_{3\vec{q}}$ are finite at $m\neq 0$,
we can replace $\left(\frac{d_{\vec{k}}+d_{3\vec{k}}}{2d_{\vec{k}} }\right)$
by its value at $\vec{k}=\vec{q}$, which is $(1+\textbf{sgn}(m))/2$ with
$\textbf{sgn}(m)$ denoting the sign of $m$.
This replacement will not
change the integral in Eq.~(\ref{eq:chernnumberA}) in the limit
$\eta \to 0$. The value of $d_{3\vec{k}}$ at $\vec{k}\neq \vec{q}$
has nothing to do with the Chern number.

From Eq.~(\ref{eq:chernnumberA}), we know that the Chern number
depends only upon $\vec{d}_{\vec{k}}$ around the singularities of $\vec{A}$. We then
expand $d_{1\vec{k}}$ and $d_{2\vec{k}}$ into power series
of $\Delta \vec{k}=\vec{k}-\vec{q}$. Without loss of generality, we have
\begin{equation}\label{eq:expdvector}
\begin{split}
d_{1\vec{k}} = & a_{1x} \Delta k_x + a_{1y} \Delta k_y + \mathcal{O}(\Delta k^2), \\
d_{2\vec{k}} = & a_{2x} \Delta k_x + a_{2y} \Delta k_y + \mathcal{O}(\Delta k^2), \\
d_{3\vec{k}} = & m + \mathcal{O}(\Delta k).
\end{split}
\end{equation}
It is straight forward to prove that the higher-order terms in this expansion do not
contribute to the integral in Eq.~(\ref{eq:chernnumberA}) in the limit
$\eta \to 0$, which evaluates
\begin{equation}\label{eq:cvecq}
 C^{(\vec{q})}= \frac{1}{2} \left( 1+\textbf{sgn}(m) \right) \textbf{sgn}
 \left( a_{1x}a_{2y}-a_{2x}a_{1y}\right).
\end{equation}
It is worth mentioning that the three components of $\vec{d}_{\vec{k}}$ are
on an equal footing. Depending on the basis that is chosen,
the components of $\vec{d}_{\vec{k}}$ could be exchanged
in some models.

Notice that, in Eq.~(\ref{eq:expdvector}), the coefficients $a_{1x}$, $a_{1y}$, $a_{2x}$
and $a_{2y}$ are $\vec{q}$-dependent.
While $m$ at different $\vec{q}_j$
may represent different parameters in the Hamiltonian,
i.e. the gap parameters at different phase boundaries.
An example is the Haldane model~\cite{supp:haldane}.
In a single Brillouin zone,
$\vec{A}$ has two singularities.
And the energy gap closes at one of them
as the system is at some phase boundary,
but closes at the other singularity
as the system is at the different phase boundary.
On the other hand, if the system has some symmetries
so that at a specific phase boundary
the gap closes simultaneously at several $\vec{q}_j$,
$m$ at these $\vec{q}_j$ must be the same parameter.

Now let us discuss the Hall conductance of quenched states
when the parameters in the initial and post-quench Hamiltonians
are both nearby a specific phase boundary where
the gap parameter is denoted by $m$.
Suppose that the system is initially in a ground state
with the gap parameter $m=m_i$, before we suddenly
change $m$ in the Hamiltonian from $m_i$ to $m_f$.
We then measure the Hall conductance in the long time limit.
The Hall conductance $C_{neq}$ is a function
of $m_f$, while we fix $m_i$ to be nonzero.

Noting $\vec{d}^i_{\vec{k}}=\vec{d}_{\vec{k}}(m_i)$
and $\vec{d}^f_{\vec{k}}=\vec{d}_{\vec{k}}(m_f)$, we express the Hall conductance as
\begin{equation}\label{eq:halltwoband}
\begin{split}
C_{neq} = \frac{1}{4\pi} \int d \vec{k}^2 & \bigg[
  \frac{ \left( \displaystyle\frac{\partial \vec{d}_{\vec{k}}(m_f)}{\partial k_x}\times
 \frac{\partial \vec{d}_{\vec{k}}(m_f)}{\partial k_y}\right) \cdot \vec{d}_{\vec{k}}(m_f)}
 {(d_{\vec{k}}(m_f))^4} \\ & \times
  \frac{\vec{d}_{\vec{k}}(m_i) \cdot \vec{d}_{\vec{k}}(m_f)
 }{{d}_{\vec{k}}(m_i) } \bigg],
 \end{split}
\end{equation}
where the integral is over a single Brillouin zone. In a generic model,
the components of $\vec{d}_{\vec{k}}$ are all analytic functions of $\vec{k}$.
According to Eq.~(\ref{eq:halltwoband}), $C_{neq}(m_f)$ is nonanalytic
only if $d_{\vec{k}}(m_f)$ in the denominator of the integral
vanishes at some $\vec{k}$, i.e., the singularities of the Berry curvature.
This is the case at $m_f=0$ when the gap of the post-quench Hamiltonian
closes at some singularities of the Berry connection $\vec{A}$.
Without loss of generality, we suppose that these singularities
are $\vec{q}_1, \vec{q}_2, \cdots, \vec{q}_{N'}$ with $N'\leq N$.
The nonanalyticity of $C_{neq}(m_f)$ comes from the integral
over the neighborhoods of $\vec{q}_1, \vec{q}_2, \cdots, \vec{q}_{N'}$.
We then divide $C_{neq}$ into the analytic
part and the nonanalytic part as we did
in the Dirac model. The latter is written as
\begin{equation}
C^\eta_{neq}= \sum_{j=1}^{N'} C_{neq}^{(\vec{q}_j)}
\end{equation}
with
\begin{equation}\label{eq:hallnonana}
\begin{split}
C_{neq}^{(\vec{q}_j)} = \frac{1}{4\pi} \int_{B_\eta(\vec{q}_j)} d\vec{k}^2 & \bigg[
  \frac{ \left( \displaystyle\frac{\partial \vec{d}_{\vec{k}}(m_f)}{\partial k_x}\times
 \frac{\partial \vec{d}_{\vec{k}}(m_f)}{\partial k_y}\right) \cdot \vec{d}_{\vec{k}}(m_f)}
 {(d_{\vec{k}}(m_f))^4} \\ & \times
  \frac{\vec{d}_{\vec{k}}(m_i) \cdot \vec{d}_{\vec{k}}(m_f)
 }{{d}_{\vec{k}}(m_i) } \bigg],
 \end{split}
\end{equation}
where $B_\eta(\vec{q}_j)$ is a circle of radius $\sqrt{\eta}$ centered at
$\vec{q}_j$ with $\eta$ a positive number that can be arbitrarily small.

In the neighborhood of the singularity $\vec{q}$, we can
expand the components of $\vec{d}_{\vec{k}}$ into power series.
Let us first consider the lowest-order term given by
Eq.~(\ref{eq:expdvector}). We substitute Eq.~(\ref{eq:expdvector}) into Eq~(\ref{eq:hallnonana}).
We replace ${d}_{\vec{k}}(m_i)$ by its value at $\vec{k}=\vec{q}$, that
is ${d}_{\vec{q}}(m_i)=|m_i|$. This replacement will not change the
nonanalytic behavior of $C_{neq}^{(\vec{q})}$ since $m_i$ is nonzero.
While the denominator of the integrand becomes
\begin{equation}\label{eq:denoexp}
 \left( d_{\vec{k}}(m_f)\right)^4 = \left( m_f^2 +
 \sum_{j=1}^2 \left( a_{jx} \Delta k_x + a_{jy}\Delta k_y \right)^2 \right)^2.
\end{equation}
We change the coordinate system so that the function $ \left( d_{\vec{k}}(m_f)\right)^4 $
has rotational symmetry around $\vec{q}$. In the new coordinate system we have
\begin{equation}
 \sum_{j=1}^2 \left( a_{jx} \Delta k_x + a_{jy}\Delta k_y \right)^2 = \Delta k'^2.
\end{equation}
This transformation is always possible. Otherwise, the coefficients before
$\Delta k_x^2$ and $\Delta k_y^2$ have different signs, which contradicts
the proposition that $\vec{q}$ is an isolated singularity. In the new
coordinate system, we carry out the azimuthal integration and obtain
\begin{equation}\label{eq:radialintexp}
\begin{split}
 C_{neq}^{(\vec{q})} = & \frac{m_f\textbf{sgn}(a_{1x}a_{2y}-a_{2x}a_{1y})}
 {4|m_i|} \\ & \times \int^\eta_0 d(\Delta k'^2) \frac{m_im_f+\Delta k'^2}{\left(
 m_f^2+ \Delta k'^2 \right)^2}.
\end{split}
 \end{equation}
In the numerator of the integrand, only the 2nd-order term $\Delta k'^2$ has a
contribution to the nonanalyticity of $C_{neq}^{(\vec{q})}(m_f)$.
It is trivial to find the original function of this integral,
whose value is an analytic function of $m_f$ at $\Delta k'^2 = \eta$
but a nonanalytic one at $\Delta k'^2 = 0$.
This coincides with our expectation that the nonanalytic
behavior of $C_{neq}^{(\vec{q})}(m_f)$ should be independent
of the choice of $\eta$. The nonanalytic part of $C_{neq}^{(\vec{q})}(m_f)$ is
\begin{equation}
 C_{neq}^{(\vec{q})} \sim \frac{-\textbf{sgn}(a_{1x}a_{2y}-a_{2x}a_{1y})}{2|m_i|}
 m_f \ln |m_f|.
\end{equation}
First, $C_{neq}^{(\vec{q})}$ is a continuous function of $m_f$,
and then the Hall conductance $C_{neq}(m_f)$ must be continuous.
Second, the derivative
of $C_{neq}^{(\vec{q})}$ with respect to $m_f$ is logarithmically divergent
in the limit $m_f\to 0$, i.e.,
\begin{equation}\label{eq:cneqqtomf}
\lim_{m_f\to 0} \frac{dC_{neq}^{(\vec{q})}}{dm_f} \sim
 \frac{-\textbf{sgn}(a_{1x}a_{2y}-a_{2x}a_{1y})}{2|m_i|}
 \ln |m_f|.
\end{equation}

Comparing Eq.~(\ref{eq:cvecq}) with Eq.~(\ref{eq:cneqqtomf}),
we immediately find that the $\vec{q}$-dependent coefficient
in $\frac{dC_{neq}^{(\vec{q})}}{dm_f}$
is equal to the change of $C^{(\vec{q})}(m_f)$ at the phase boundary $m_f=0$.
$C_{neq}^\eta$ is the sum of $C_{neq}^{(\vec{q}_j)}$
at the singularities $\vec{q}_1, \vec{q}_2, \cdots, \vec{q}_{N'}$,
while the Chern number $C$ is the sum of $C^{(\vec{q}_j)}$
at all the singularities of $\vec{A}$. But
$C^{(\vec{q}_j)}$ at $j>N'$ does not change at $m_f=0$, since the
corresponding gap parameter is different from $m_f$.
We finally obtain
\begin{equation}\label{eq:supp:central}
 \lim_{m_f\to 0} \frac{dC_{neq}}{dm_f} \sim
 \frac{\displaystyle\lim_{m_f\to 0^-} C(m_f)- \displaystyle\lim_{m_f\to 0^+} C(m_f) }{2|m_i|}
 \ln |m_f|,
\end{equation}
which is the central result of this paper.

Eq.~(\ref{eq:supp:central}) is obtained by considering only
the lowest-order term in the expansion of $\vec{d}_{\vec{k}}$.
Next we prove that the higher-order terms do not change the continuity of
$C_{neq}$ or the asymptotic behavior
of $dC_{neq}/dm_f$ in the limit $m_f\to 0$. This is true if the higher-order
terms do not change the continuity of
$C_{neq}^{(\vec{q})}$ or the asymptotic behavior
of $dC_{neq}^{(\vec{q})}/dm_f$ at an arbitrary singularity.

A linear term like
$\left(a_{3x} \Delta k_x+a_{3y} \Delta k_y\right)$ is not allowed in
the expansion of $d_{3\vec{k}}$ in Eq.~(\ref{eq:expdvector}).
Otherwise, $d_{\vec{q}}$ is not the energy gap, or the
minimum point of $d_{\vec{k}}$ is not at $\vec{k}=\vec{q}$,
but changes with $m$, which contradicts our proposition.
In a generic model like the Dirac model,
the Haldane model or the Kitaev honeycomb model, the minimum point of $d_{\vec{k}}$
is determined by the symmetry of the model and then keeps invariant
as the system is in the vicinity of the phase boundary.

Let us add the 2nd-order term into $d_{3\vec{k}}$, i.e.,
$\left(b_{3x}\Delta k_x^2 + b_{3y}\Delta k_y^2 + b_{3m} \Delta k_x \Delta k_y \right)$
without loss of generality.
The denominator in the integrand of $C_{neq}^{(\vec{q})}$
becomes
\begin{equation}\label{eq:cneqdenohigh}
\begin{split}
 \left( d_{\vec{k}}(m_f)\right)^4 = & \bigg[ m_f^2 +
 \sum_{j=1}^2 \left( a_{jx} \Delta k_x + a_{jy}\Delta k_y \right)^2 \\ &
 + 2m_f\left(b_{3x}\Delta k_x^2 + b_{3y}\Delta k_y^2 + b_{3m} \Delta k_x \Delta k_y \right)
\\ &  + \mathcal{O}(\Delta k^4) \bigg]^2.
 \end{split}
\end{equation}
$C_{neq}^{(\vec{q})}$ is an integral over the infinitesimal neighborhood
of $\vec{q}$, where the 4th-order term $ \mathcal{O}(\Delta k^4)$
is much smaller than the 2nd-order term and can be neglected.
At the same time, the additional 2nd-order term that is proportional to $m_f$
has no contribution to the asymptotic behavior of $C_{neq}^{(\vec{q})}$ and $dC_{neq}^{(\vec{q})}/dm_f$
in the limit $m_f\to 0$. Therefore, the effective denominator
is the same as Eq.~(\ref{eq:denoexp}). The numerator of the integrand
becomes
\begin{equation}
 \begin{split}
\bigg[ & \left( \displaystyle\frac{\partial \vec{d}_{\vec{k}}(m_f)}{\partial k_x}\times
 \frac{\partial \vec{d}_{\vec{k}}(m_f)}{\partial k_y}\right) \cdot \vec{d}_{\vec{k}}(m_f)\bigg]
\left(\vec{d}_{\vec{k}}(m_i) \cdot \vec{d}_{\vec{k}}(m_f)\right) \\
 = & (a_{1x}a_{2y}-a_{2x}a_{1y}) \bigg[ m_im_f^2+m_f \sum_{j=1}^2 \left(a_{jx}\Delta k_x
 +a_{jy}\Delta k_y\right)^2 \\ & + m_f^2 \left(b_{3x}\Delta k_x^2+
b_{3y}\Delta k_y^2+ b_{3m}\Delta k_x \Delta k_y\right) + \mathcal{O}(\Delta k^4) \bigg].
 \end{split}
\end{equation}
The 4th-order term $\mathcal{O}(\Delta k^4)$ can be neglected in the limit $\eta \to 0$.
This can be easily
verified by adding $\Delta k'^4$ in the numerator of the integrand in Eq.~(\ref{eq:radialintexp})
and checking the output. The additional 2nd-order term that is proportional
to $m_f^2$ leads to a correction of $C_{neq}^{(\vec{q})}\sim m_f^2 \ln |m_f|$, which
does not change the asymptotic behavior of $C_{neq}^{(\vec{q})}$ and $dC_{neq}^{(\vec{q})}/dm_f$
in the limit $m_f\to 0$.
In the power series of $d_{3\vec{k}}$, any term in order higher than $2$
leads to a correction to numerator or denominator of the integrand which is
at least in the 3rd order of $\Delta k$ and can then be neglected in the limit $\eta \to 0$.
Therefore, the higher-order terms in $d_{3\vec{k}}$ do not affect the
asymptotic behavior of $dC^{(\vec{q})}_{neq}/dm_f$ or the continuity
of $C^{(\vec{q})}_{neq}$.

Similarly, we can prove that the higher-order terms in $d_{1\vec{k}}$
or $d_{2\vec{k}}$ have no contribution. In fact, the terms in order higher than
$1$ lead to a correction of $\mathcal{O}(\Delta k^3)$ in the denominator.
The terms in order higher than $3$ also
lead to a correction of $\mathcal{O}(\Delta k^3)$ in the numerator,
which can be neglected. The 2nd- and 3rd-order terms in $d_{1\vec{k}}$
or $d_{2\vec{k}}$ generate a linear term and a 2nd-order term that
is proportional to $m_f^2$ in the numerator. The latter
does not contribute to the
asymptotic behavior of $dC^{(\vec{q})}_{neq}/dm_f$ due to the similar reason mentioned above.
While the linear term in the numerator is an odd function of $\Delta k_x$ or $\Delta k_y$,
and then has no contribution to the integral since both
the denominator and the integration boundary
have rotational symmetry with respect to the singularity.


\begin{references}
\bibitem{universalargue} These results will be published separately.
\end{references}
\begin{references}
\bibitem{continuity} In this paper, ``$C_{neq}(M_f)$ is continuous'' means
$\left( \lim_{M_f \to 0^+}C_{neq} = \lim_{M_f \to 0^-}C_{neq} \right)$. Notice that these limits
are not necessarily equal to $C_{neq}(M_f=0)$.
\end{references}
\end{document}